\documentclass{article}

\usepackage{arxiv}

\usepackage[utf8]{inputenc} 
\usepackage[T1]{fontenc}    
\usepackage{hyperref}       
\usepackage{url}            
\usepackage{booktabs}       
\usepackage{amsfonts}       
\usepackage{nicefrac}       
\usepackage{microtype}      
\usepackage{makecell}
\usepackage{siunitx}
\usepackage{caption}
\usepackage{subcaption}
\usepackage{multirow}
\usepackage{xr}
\usepackage{cleveref}
\usepackage{graphicx}
\usepackage{epsfig}
\usepackage{tikz}

\pdfoutput=1

\title{Emergent Influence Networks in Good-Faith Online Discussions}

\author{
    Henry K. Dambanemuya \\
  Northwestern University\\
  Evanston, IL 60208 \\
  \texttt{hdambane@u.northwestern.edu} \\
   \And
  Daniel Romero \\
  University of Michigan \\
  Ann Arbor, MI 48109 \\
  \texttt{drom@umich.edu} \\
  \And
  Em\H oke-\'Agnes Horv\'at \\
  Northwestern University\\
  Evanston, IL 60208 \\
  \texttt{a-horvat@northwestern.edu } \\
}

\begin{document}
\maketitle

\begin{abstract}
Town hall-type debates are increasingly moving online, irrevocably transforming public discourse. Yet, we know relatively little about crucial social dynamics that determine which arguments are more likely to be successful. This study investigates the impact of one's position in the discussion network created via responses to others' arguments on one's persuasiveness in unfacilitated online debates. We propose a novel framework for measuring the impact of network position on persuasiveness, using a combination of social network analysis and machine learning. Complementing existing studies investigating the effect of linguistic aspects on persuasiveness, we show that the user's position in a discussion network influences their persuasiveness online. Moreover, the recognition of successful persuasion further increases this dominant network position. Our findings offer important insights into the complex social dynamics of online discourse and provide practical insights for organizations and individuals seeking to understand the interplay between influential positions in a discussion network and persuasive strategies in digital spaces.
\end{abstract}

\keywords{persuasion, communication networks, online communities}

\subsection{Introduction}

Persuasion is the process of influencing or changing one's opinions, beliefs, or behavior through argumentation~\cite{gass2010persuasion,cialdini1993influence}. It is critical in advertising, marketing, political campaigns, and interpersonal communication~\cite{shrum2012persuasion,fogg2002persuasive}. Theories of persuasion, e.g., the Elaboration Likelihood Model (ELM), suggest that people differ in the extent to which they are willing to engage with an argument  thoughtfully~\cite{petty1986elaboration}. Those with high levels of elaboration are more likely to scrutinize the merits of the arguments presented to them. However, those with low levels of elaboration are more likely to consider peripheral factors, such as specific argument-independent characteristics of the speaker. Extensive literature in social psychology and sociology suggests that a speaker's position in a communication network is linked to how influential they are in changing others' opinions in political or health-related settings~\cite{centola2018truth,centola2013social,christakis2007spread,christakis2008collective}. 

Many public town hall-type debates have moved to digital spaces in recent decades. This shift has fundamentally transformed the nature of public discourse. For example, online social networks have made it easier for people to discuss with others, regardless of their physical location. Additionally, online spaces promise fewer gatekeepers and less salient signals of power and stature than in-person meetings. Thus, while the transition to online spaces could potentially lead to a diversification of voices and perspectives in public conversations, it raises questions about the quality of discourse and the impact of network dynamics on persuasion even without clearly signaled hierarchies~\cite{stromer2003diversity}. In other words, do substantive arguments or the user's network position determine which arguments will be more persuasive? To answer this question, here \textit{we empirically investigate the role of network position in persuasiveness online}. 

The complex communication, persuasion, and decision-making dynamics in contemporary society have long been popular research subjects~\cite{christakis2009connected,fogg2002persuasive,tan2016winning}. In particular, the evolving structures of connections among individuals (or organizations) that shape the flow of information, ideas, and resources are fundamental for the critical social processes that are at play during discussions~\cite{borgatti2011network}. These networks frequently encode influence in a subtle way. Since they are characterized by fluidity, adaptability, and responsiveness to changing contexts~\cite{cross2004hidden}, they are typically difficult to map and study at scale.

The rise of social media platforms, such as Facebook, Twitter, Instagram, and Reddit has facilitated the development of influence networks, all the while creating the digital footprints necessary for examining them systematically~\cite{lutu2019using,wu2015analyzing}. Using data on who responds to whom, how and with what effect, we create discussion networks. Based on these networks, our work can move beyond existing efforts to study prominent ``influencers,'' i.e., the few individuals or organizations with the ability to shape the opinions, attitudes, and behaviors of their followers through their curated and highly visible activities~\cite{panagopoulos2020influence}. Our research scrutinizes thus the conditions under which argumentation can be effective without the ``social clout''  of high-profile influencers~\cite{lutu2019using} and can be generalized to the majority of online platform users.

Studying persuasiveness on social media platforms and fora requires a complex systems view, as these networks represent emergent systems that cannot be understood by only analyzing their individual components~\cite{conte2012manifesto}. Persuasion is often a collective rather than an individual effort even in unfacilitated discussions without strict external moderation that guides the discussion, enforcing community governing principles. Instead, many individuals' contributions lead to the argumentation's success or failure in a self-organizing process. For this reason, we consider social interactions among the participants in online discourses that do not feature outside interventions but are still subject to the emergence of differences in people's influence that have a non-negligible impact on persuasion~\cite{zeng2020changed,tan2016winning}.

Within this approach, our study investigates whether, in addition to the linguistic quality of their argument, one's position in the discussion network (i.e., being influential or not) plays a role in their persuasiveness. To better understand the link between emergent influence and persuasion power, we examine successful arguments' effect on one's network position. Does successful persuasion have a positive reinforcing impact on one's influence? Or is influence a fleeting asset in these otherwise unstructured discussions? This inquiry is critical to uncover the potential endurance and implications of emergent influence networks in online discussions. Our research focuses on a platform that hosts good-faith discussions on Reddit. This choice was motivated by extensive NLP-based work that uncovered the linguistic factors determining which arguments were persuasive~\cite{tan2016winning,khazaei2017writing}. We use this prior research as a baseline to understand the role of networks above and beyond crucial language markers.
Finally, we discuss the implications of our findings in eliciting opinion change related to real-world problems on this platform, such as sustainability, migration, and global health.

Investigations into the role of one's network position in online discussions are important because they can shed light on how emergent influence networks determine the persuasiveness of information and ideas. This work can also help us understand how arguments are received and evaluated in online communities and how the structure and dynamics of discussion networks shape people's attitudes and beliefs. Furthermore, our research has implications for the design of online communication platforms and the development of strategies to enhance the quality of online discourse.

\section{Related Work}

\subsection{Theories of persuasion}

To investigate the role of different factors in determining persuasiveness, we rely on the classical \emph{Elaboration Likelihood Model (ELM)} as a theoretical base. The ELM provides a framework for understanding the basic processes underlying the effectiveness of persuasion and attitude change~\cite{petty1986elaboration}. The theory proposes two different processes by which people can be persuaded: a \emph{central route} and a \emph{peripheral route}. The central route focuses on a message's quality to influence opinions. For instance, the argument's readability and quality are perceived via a central route by audiences who carefully weigh its merits~\cite{o1995argument,khazaei2017writing}. The peripheral route occurs when a message receiver is unable or unwilling to decode the argument thoughtfully and instead relies on positive or negative cues associated with the message (e.g., the credibility or attractiveness of the sources of the message)~\cite{han2018persuasion}. 

\begin{figure}[!h]
    \centering
    \includegraphics[scale=.4]{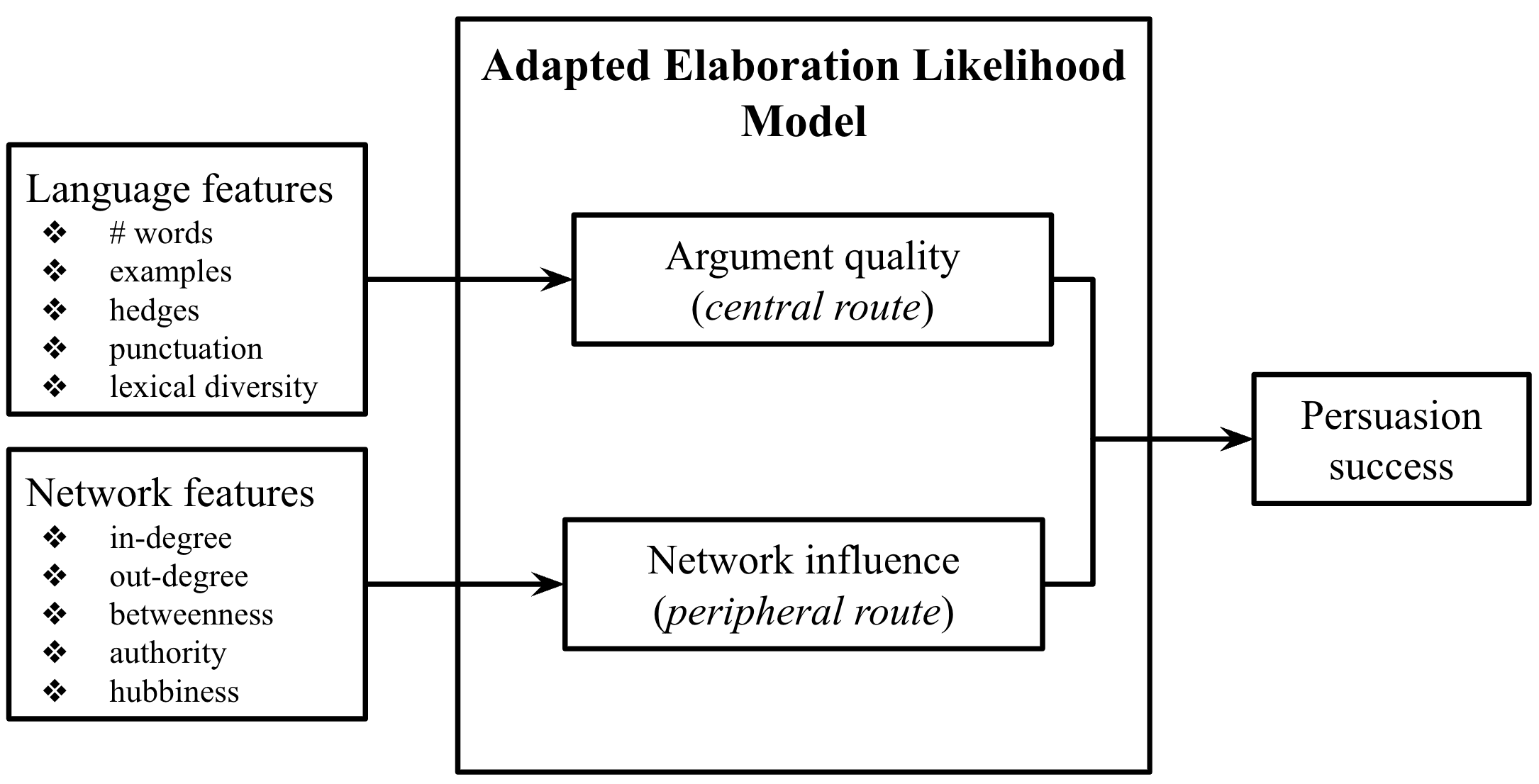}
    \caption{Proposed adaptation of the Elaboration Likelihood Model (ELM) with language features as the central route of persuasion and network features as the peripheral route.}
    \label{fig:adaptedmodel}
\end{figure}

Similarly to ELM's peripheral route, the \emph{Pre-suasion Model} of persuasion also suggests a way to influence others by channeling attention to the persuader before exposure to the persuasive message~\cite{cialdini2016pre}. For example, individuals on online platforms may first seek to build trust, credibility, or popularity with their audience before attempting to persuade them. For example, high popularity can pre-suade audiences by shifting their attention to the popular individual's argument.
Adapting these theories to the setting of online discussions, we identify from existing literature a set of factors related to the central and peripheral routes of persuasion, as shown in Figure~\ref{fig:adaptedmodel} and detailed next.

\subsection{The language of persuasion}

Extensive empirical research investigates linguistic factors associated with persuasive messages, including vocabulary and topic extraction \cite{althoff2014ask,tan2016winning}, the use of semantic and syntactic rules \cite{hidey2017analyzing}, and argument interactions and structure \cite{ji2018incorporating, wei2016post, tan2016winning}. These language factors are essential for the central route through which people process the persuasiveness of a message. The language of persuasion has been studied in text-based contexts online, including on crowdfunding platforms, advertising and marketing sites, product recommendations, and design practices that promote behavioral change~\cite{mitra2014language,shrum2012persuasion,li2011online,fogg2002persuasive}. 

Crowdfunding studies investigate linguistic factors associated with successful fundraising for lending~\cite{larrimore2011peer,han2018persuasion}, patronage~\cite{mitra2014language}, and charitable~\cite{rhue_emotional_2018,liang2016inspire} campaigns.
For example, Larrimore et al~\cite{larrimore2011peer} rely on the Linguistic Inquiry and Word Count (LIWC) software to examine the relationship between language use and persuasiveness in peer-to-peer lending. They find that the use of extended narratives, concrete descriptions and quantitative words that are likely related to one's financial situation are positively associated with successfully receiving a loan. 
Focusing on patronage sites such as KickStarter.com, Mitra and Gilbert~\cite{mitra2014language} also find persuasive language factors that predict funding success, e.g., phrases that signal lucrative offers or the attention that a project has already received. 
In the lack of financial or material incentives, i.e., in prosocial fundraising, Rhue and Robert~\cite{rhue_emotional_2018} find that using emotions can effectively persuade others to contribute to one's campaigns. 
Other studies of altruistic requests in online communities also show that communicating one's needs clearly and including linguistic indications of gratitude and evidentiality are essential to convince donors to contribute to philanthropic causes~\cite{althoff2014ask}. 

Recently, the subreddit community \texttt{r/ChangeMyView} received significant research attention~\cite{tan2016winning,hidey2017analyzing,wei2016post}. For example, Tan et al!\cite{tan2016winning} examined the similarity between the language of the opinion holder (the person requesting arguments to refute their initial statement) and the counterarguments (the responses that challenge the initial opinion). They find that a high linguistic similarity is predictive of persuasiveness. Hidey et al~\cite{hidey2017analyzing} annotated different types of semantic claims (e.g., interpretation, evaluation, agreement, and disagreement) to investigate whether certain claims are more likely to appear in persuasive than non-persuasive messages. The authors find that agreeing with what was previously said by others before expressing a diverging opinion, conceptual coherence (i.e., consecutive arguments of the same type), and claims backed by premises constitute persuasive rhetorical strategies.

\subsection{Persuasion and networks}

As opposed to ample research on linguistic factors associated with persuasion, far less work focuses on other factors that may initially be processed through the peripheral routes of information processing but become critical for persuasive communication. Among these peripheral features, critical factors are related to people's influence in the discussion networks.

Social networks play a significant role in persuasion due to their influence on individuals' attitudes, beliefs, and behaviors~\cite{christakis2009connected,centola2018truth}. Extant studies suggest that social network structure determines the impact of social influence, especially when it comes to the large-scale adoption of health-related~\cite{christakis2007spread,christakis2008collective,centola2013social}, community-building~\cite{blair2019motivating}, and sustainable behaviors~\cite{constantino2022scaling,white2019shift}. The success of persuasion via social influence depends on the nature of information flows. Specifically, simple vs complex contagion can predict the successful adoption of ideas in social networks~\cite{centola2018truth,guilbeault2018complex}. Whereas simple contagions (e.g., the spread of news much like the transmission of the flu) require a single contact activation, the complex persuasive ideas and behavioral change require complex contagions, i.e., reinforcement of information that results from multiple sources of interaction~\cite{centola2018truth,guilbeault2018complex}. Thus, networks play a crucial role in understanding persuasiveness in online settings, as network structures that facilitate simple contagion can hinder complex contagion processes that are crucial for persuasion success.

\section{Data}

We rely on data from the subreddit \texttt{r/ChangeMyView}, an active community on Reddit that started in January 2013 as a forum for debate and has grown to over 3 million users as of March 2023. \texttt{r/ChangeMyView} provides a platform for people to discuss a wide range of topics (e.g., abortion, gun control, vaccinations, taxes, feminism, marriage, religious freedom, climate change, and society) to understand opposing viewpoints. On the platform, an original poster (OP) begins by posting an opinion or belief they hold to be accurate but accept that it may be flawed. They also share their reasoning behind their opinion in at least 500 characters. The OP is interested in understanding other perspectives on the issue, so challengers are invited to contest the OP's viewpoint. OPs explicitly recognize challengers' successful arguments by replying with the $\Delta$ character and explaining how and why their original viewpoint changed. The specific data sample we analyse here was released by Tan et al~\cite{tan2016winning} and consists of discussions from January 2013 to May 2015.

Using basic natural language processing (NLP), we follow each discussion's comment threads to identify persuasive challengers awarded $\Delta$s. The community also has strict rules to facilitate good-faith discussions. For example, the OP must personally hold the view they are offering for discussion, demonstrate that they are open to it changing, and be willing to have a conversation within 3 hours after posting. Challengers' direct responses to an OP must question at least one aspect of the OP's viewpoint, contribute meaningfully to the discussion and not be rude or hostile to other users.

Our data includes the full text of arguments and the structure of responses (i.e., ``in-reply-to'' relationships between arguments). These two types of information enable us to deduce various language and network features, which could be correlated with persuasiveness. In what follows, we describe both language and network features.

\section{Measures}

\subsection{Language features}

We adopt a set of measures that have been shown to be associated with persuasiveness in the \texttt{r/ChangeMyView} community~\cite{tan2016winning,khazaei2017writing}. The used language features encompass the number of (in)definite articles, which are associated with an argument's specificity~\cite{danescu2012you}; positive and negative words suggestive of patterns of emotion~\cite{hullett2005impact,wegener1996effects}; question and quotation marks prompting clarification and indicating attention to others' words~\cite{khazaei2017writing}; personal pronouns with self-affirmation and examples from personal experiences in argumentation~\cite{correll2004affirmed,cohen2000beliefs}; and the number of URL links with citing external evidence to support an argument~\cite{tan2016winning}. We consider the number of words, sentences, and the readability of the arguments quantified by the Flesch-Kincaid grade level and readability score~\cite{flesch2007flesch}. We include lexical diversity via a word entropy, which refers to the Shannon entropy of the set of words in the argument, and a token type entropy meaning the Shannon entropy of the set of different parts of speech used in the argument. Finally, we rely on a set of features that describe an argument's (A) vocabulary overlap with the original post (O) using the following measures~\cite{tan2016winning}: 

\begin{itemize}
    \item \emph{Number of common words} between an argument and original post: $|A \bigcap O|$
    \item \emph{Reply fraction} of common words in the argument: $\frac{|A \bigcap O|}{|A|}$
    \item \emph{OP fraction} of common words in the original post: $\frac{|A \bigcap O|}{|O|}$
    \item \emph{Jaccard similarity} between the original post and the argument: $\frac{|A \bigcap O|}{|A \bigcup O|}$
\end{itemize}

Following Tan et al~\cite{tan2016winning}, we compute the language features for each persuasive challenger who received a $\Delta$ and a matched challenger who was not awarded a $\Delta$ but had the highest overlap in their arguments' vocabulary with the successful challenger. This matched sample allows us to compare properly challengers with lexically similar arguments.

\subsection{Network Features} 

To explore the structure of discussions, we create directed discussion networks whose nodes are \texttt{r/ChangeMyView} users (i.e., the OP and their challengers) and whose edges denote ``in-reply-to'' relations arising from a challenger replying to another challenger or the OP. Notice that this network is weighted because, for instance, the same challenger can provide arguments to the OP multiple times during the discussion.

We calculate each node's centrality in these discussion networks to quantify the user's influence based on their network position. We choose five types of centralities, which are based purely on structural information as follows:

\begin{itemize}
    \item \textbf{In-degree}. This centrality is defined as the sum of all incoming edge weights such that the measure quantifies the number of replies the challenger received from other challengers or the OP. In-degree has been used to quantify the popularity of social network users on platforms such as Twitter, Instagram, and Facebook~\cite{lutu2019using}. Typically, individuals with high in-degree centrality occupy an advantageous social position because they have more direct sources of information~\cite{wasserman1994social,johnson2018analyzing}.
    \item \textbf{Out-degree}. We sum over the weights of all outgoing edges from each node to obtain a node's out-degree. This centrality quantifies how often the challenger replied to others. Proactively connecting arguments or mediating in tense discussions might indicate a critical network position.
    \item \textbf{Degree ratio}. As a measure of balance between incoming and outgoing edge weights, we compute the degree ratio as the ratio between out-degree and in-degree (i.e., out-degree/in-degree).
    \item \textbf{Authority}. This centrality is based on the Hyperlink-Induced Topic Search (HITS) algorithm developed by Kleinberg~\cite{kleinberg1998authoritative}. It quantifies the number and quality of links pointing \textit{to} the challenger from other high-authority users in the discussion network. It has been shown to be associated with the number of positive evaluations in crowd innovation contests that collect novel ideas from consumers~\cite{ozaygen2018idea}.
    \item \textbf{Hubbiness}. This centrality is also based on the HITS algorithm~\cite{kleinberg1998authoritative}, but it quantifies the number and quality of links pointing \textit{from} the challenger to other high-authority nodes in the network. It has been found to be correlated with enhanced learning from various network communities~\cite{taylor2022building}.
    \item \textbf{Betweenness}. This centrality quantifies the probability that a node lies on the shortest (directed) path between two randomly chosen nodes. It has been used to identify influential nodes or opinion leaders in online social networks~\cite{opuszko2019peer,johnson2018analyzing}.
\end{itemize}

\begin{figure}[!h]
    \centering
    \includegraphics[scale=.4]{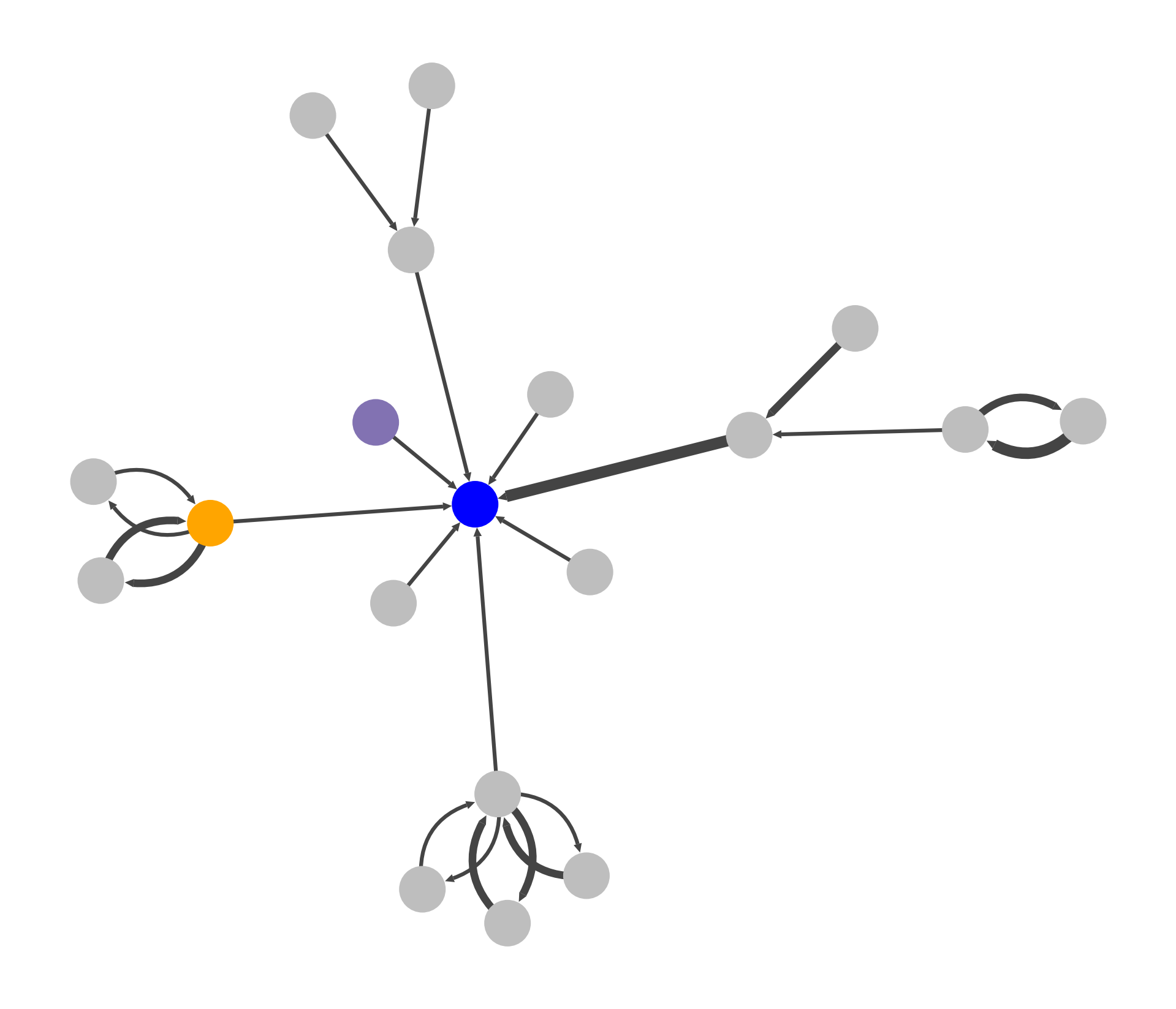}
    \includegraphics[scale=.4]{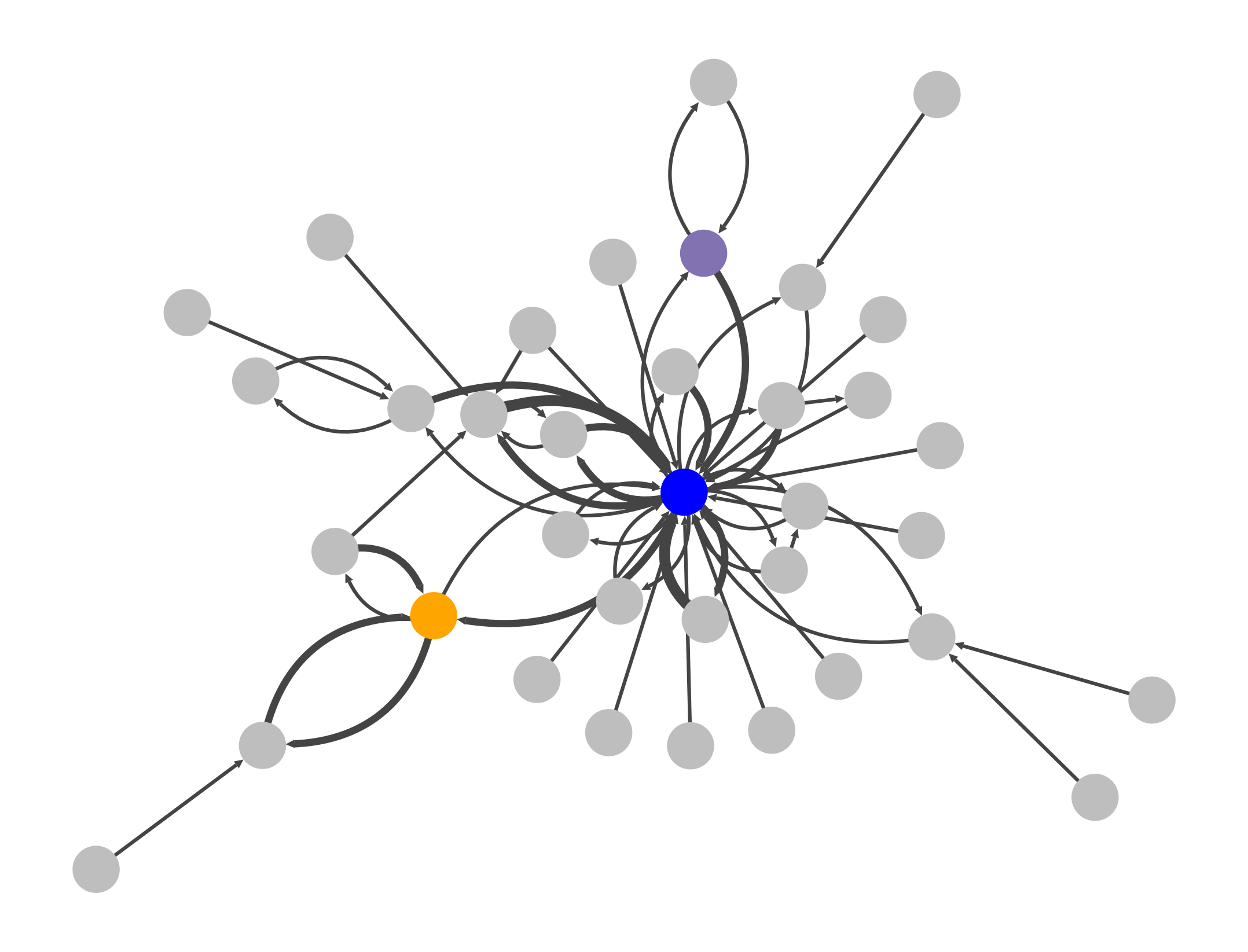}
    \caption{Example illustration of a discussion network immediately before an OP awards a $\Delta$ (left) and when the conversation concludes at a later time (right). The node represent the OP (blue), successful challenger (orange), matched unsuccessful challenger (purple), and other unsuccessful challengers (grey). The directed edges represent who replied to whom. The edge weights denote the number of replies and are signaled by line thickness.}
    \label{fig:network_illustration}
\end{figure}

These centralities inherently change throughout the evolution of the discussion. We compute them right before a $\Delta$ is awarded to reflect the influence accumulated by the successful challenger at that moment. For proper comparison, we compute the centralities for a matching challenger with similar arguments at the same time in the discussion. See Figure~\ref{fig:network_illustration} for an example network illustration before and after a $\Delta$ is awarded by an OP.

\section{Methods}

\paragraph{Identifying features associated with persuasiveness.} We train and evaluate the performance of supervised classification models to predict which challengers will successfully persuade an OP to change their view. Thus, the outcome of a classification is whether or not a challenger's arguments are persuasive. If the challenger is awarded a $\Delta$, the outcome is 1; otherwise, it is 0. We perform the classification based on the challenger's language and network features computed on observations immediately before the OP awards a $\Delta$. To ensure generalizability, we use various models, including Decision Trees, Random Forests~\cite{breiman2001random}, Adaptive Boosting~\cite{freund1997decision}, Logistic Regression, and Gaussian Naive Bayes. We report the area under the receiver operating characteristic curve (AUC) scores computed on hold-out test sets during 5-fold cross-validation to evaluate classification model performance. We rely on Random Forest permutation importance scores to evaluate the relative importance of network features and select the network feature with the most explained variance in predicting a challenger's persuasiveness.

\paragraph{Difference in Difference (DID) estimation in the change of network centrality over time.} To estimate the effect of receiving a $\Delta$ on users' network position, we rely on an econometric approach for inferring the causal effects of treatment. In other words, we compare longitudinal data between a treatment group (users awarded $\Delta$s) and a control group of matching users not awarded $\Delta$s. Specifically, we compare the magnitude of the gap between the treatment and control groups \textit{before} and \textit{after} an OP awards a $\Delta$ to a challenger. We estimate the DID effect using the following specification: $ y = \beta_0 + \beta_1T + \beta_{2}G + \beta_3(T \cdot G) + \epsilon$, where $y$ is a type of network centrality, $B_0$ is the regression intercept, $G$ is a dummy variable for group membership (0 = control, 1 = treated), $T$ is a dummy variable for the period (0 = before the $\Delta$, 1 = after the $\Delta$), $T \cdot G$ is the interaction term between time and treatment group, and $\epsilon$ is the error term. The coefficient $\beta_3$ of the interaction of $T$ and $G$ is the DID effect. 

\section{Results}

We analyzed 283,751 comments made by 34,907 challengers in 3,051 posts. On average, each post contains 93 comments and 38 unique challengers. Of the 3,051 posts, 1,741 (57.06\%) posts had OPs who changed their viewpoints. On average, posts with opinion change receive 58.43 (standard deviation = 98.54) replies before a challenger receives a $\Delta$ and 33.88 (standard deviation = 81.30) replies after a $\Delta$ is awarded. Thus, conversations in the \texttt{r/ChangeMyView} persists despite OPs acknowledging when they are willing to change their original viewpoint. Among the 1,741 posts in which opinion change occurred, we further identified 3,098 (1.09\%) persuasive comments from 1,503 (4.31\%) challengers that were awarded $\Delta$s by an OP. An additional 776 (0.27\%)  comments from 652 (1.87\%) challengers were awarded $\Delta$s by other challengers in the conversations.

\paragraph{Language Features.} Consistent with prior findings, we observe that successful arguments are more wordy, use more URL links as supporting evidence, utilize more examples (i.e., phrases such as ``for example'', ``for instance'', or ``e.g.'') and punctuation, have more lexical diversity (e.g., as measured by different words or different parts of speech), and have more words in common with the original post compared to unsuccessful arguments, c.f., ~\cite{tan2016winning,khazaei2017writing}. We also find that successful arguments contain both more positive and negative words thereby supporting two contrasting hypotheses in the existing literature. On the one hand, existing literature suggests that the use of positive words may lead to persuasion success by energizing and directing one's behavior to react positively to an argument or request~\cite{liang2016inspire}. On the other hand, the ``empathy helping hypothesis'' suggests that the use of negative words can lead to persuasion success by making people more empathetic towards one's plight~\cite{fisher2008empathy}. Determining types of arguments in which either positive or negative words will result in persuasion success is beyond the scope of our work. Table~\ref{tab:summary} provides a comparison between (un)successful arguments.

\begin{table}[!h]
\resizebox{\columnwidth}{!}{
\begin{tabular}{lll}
\hline
\begin{tabular}[c]{@{}l@{}}Features \\ \end{tabular} & \begin{tabular}[c]{@{}l@{}}Non-Persuasive \\ Challengers\end{tabular} & \begin{tabular}[c]{@{}l@{}}Persuasive \\ Challengers\end{tabular} \\ \hline
\textit{Language features} &  &  \\
\# words & 305.318 (297.102) & 440.757 (432.011) \\
\# sentences & 17.052 (17.730) & 23.476 (22.895) \\
\# positive words & 8.612 (9.254) & 12.420 (13.793) \\
\# negative words & 6.859 (9.260) & 9.876 (12.411) \\
\# examples & 1.295 (2.461) & 1.892 (3.102) \\
\# hedges & 5.740 (6.498) & 8.017 (8.898) \\
\# definite articles & 6.068 (7.211) & 8.883 (9.928) \\
\# indefinite articles & 0.091 (0.327) & 0.116 (0.350) \\
\# 1st person pronouns & 5.891 (9.301) & 8.003 (11.227) \\
\# 1st person plural pronouns & 1.692 (3.417) & 2.614 (5.119) \\
\# question marks & 1.860 (3.309) & 2.304 (3.649) \\
\# quotation marks & 11.274 (13.611) & 16.007 (19.065) \\
\# url & 0.076 (0.535) & 0.123 (0.653) \\
word entropy & 6.672 (0.792) & 7.018 (0.700) \\
token type entropy & 3.534 (0.175) & 3.565 (0.099) \\ 
\# common words & 44.298 (28.431) & 50.422 (30.736) \\
reply fraction & 0.285 (0.110) & 0.245 (0.103) \\
OP fraction & 0.241 (0.125) & 0.271 (0.128) \\
Jaccard similarity & 0.132 (0.051) & 0.129 (0.045) \\ \hline
\textit{Network features} &  &  \\
in-degree & 1.515 (3.171) & 1.374 (2.909) \\
out-degree & 2.015 (2.771) & 2.140 (2.524) \\
degree ratio & 1.508 (0.576) & 1.664 (0.548) \\
betweenness & 20.971 (235.408) & 27.275 (222.365) \\
authority & 0.039 (0.140) & 0.036 (0.135) \\
hubbiness & 0.492 (0.320) & 0.523 (0.328) \\ \hline
\end{tabular}
}
\caption{Mean (standard deviation) statistics of language and network features for all matched user pairs computed immediately before an OP awards a $\Delta$ point.}
\label{tab:summary}
\end{table}

\begin{figure}[!h]
    \centering
    \includegraphics[scale=.75]{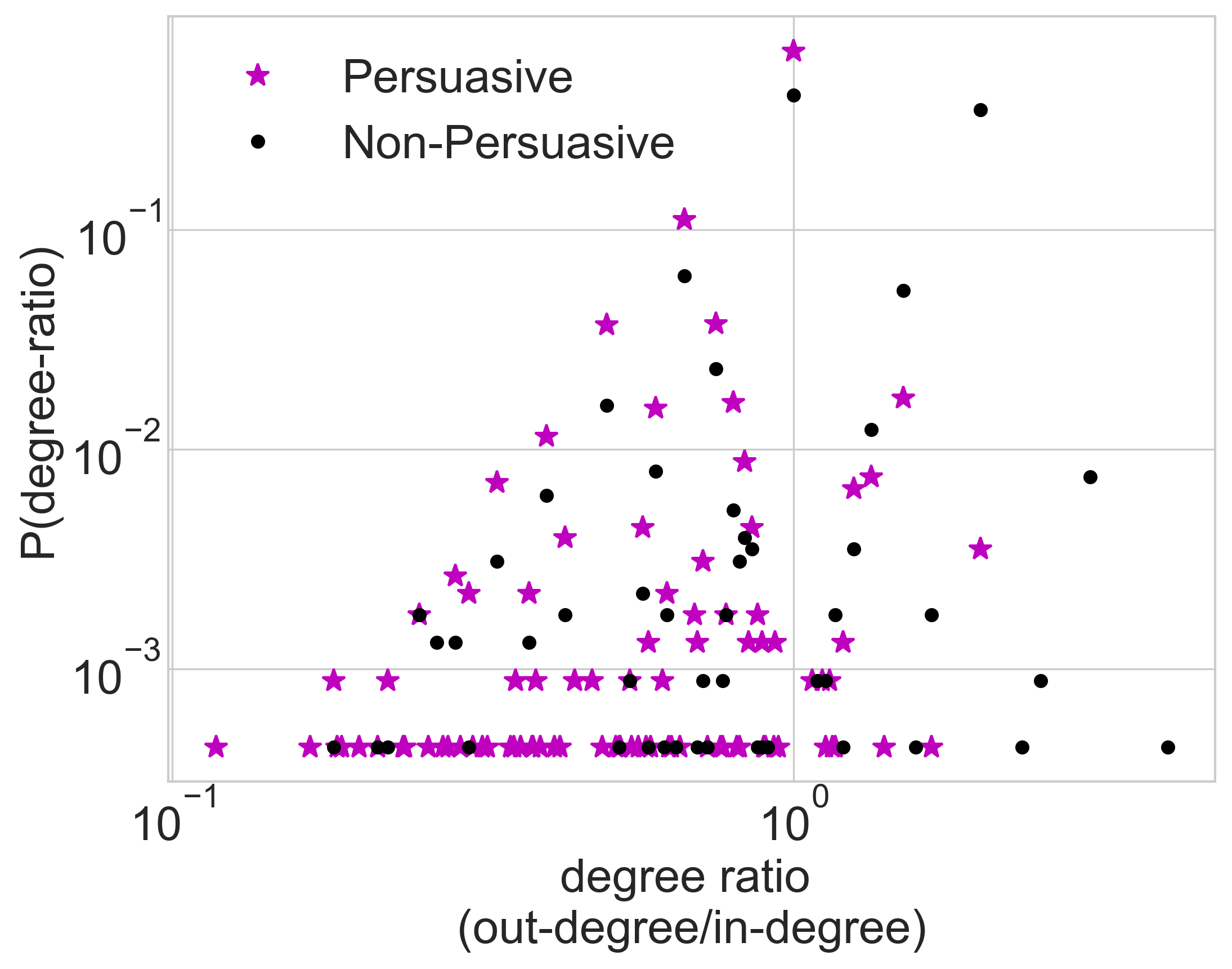}
    \caption{Degree-ratio distributions for (non)-persuasive arguments.}
    \label{fig:nx_distributions}
\end{figure}

\paragraph{Network Features.} We observe that having higher initial out-degree, hubbinness, and betweeness centralities is positively associated with successful argumentation (Table~\ref{tab:summary}). Accordingly, the more arguments one provides (hence higher out-degree), the more likely that they will succeed. The higher the number and quality of links pointing from the challenger to other high-authority challengers (hence greater hubbiness), the more likely a challenger will succeed. Additionally, the observation that higher betweenness centrality is associated with persuasion success implies that challengers who play a crucial role in transmitting information between different parts of the communication network have better chances of persuasion success. 

However, having higher initial in-degree and authority centrality is negatively associated with successful argumentation (Table~\ref{tab:summary}). Thus, prior to getting recognition from an OP, we observe that successful challengers reach out to other challengers more than they receive replies and hence they also have a higher degree ratio than unsuccessful challengers (Figure~\ref{fig:nx_distributions}). This finding suggests an ``exploration $\rightarrow$ exploitation'' argumentation strategy whereby successful challengers might first explore multiple viewpoints that enable them to combine different viewpoints and then provide substantive arguments (c.f., Figure~\ref{fig:network_illustration}). 

We further observe that most network centralities are highly correlated. Using Random Forest permutation importance score, we observe that degree ratio has the highest importance, explaining 32.9\% of the variance among network features.

\subsection{Network features enhance the prediction of persuasion}

\begin{figure}[!h]
    \centering
    \includegraphics[scale=.5]{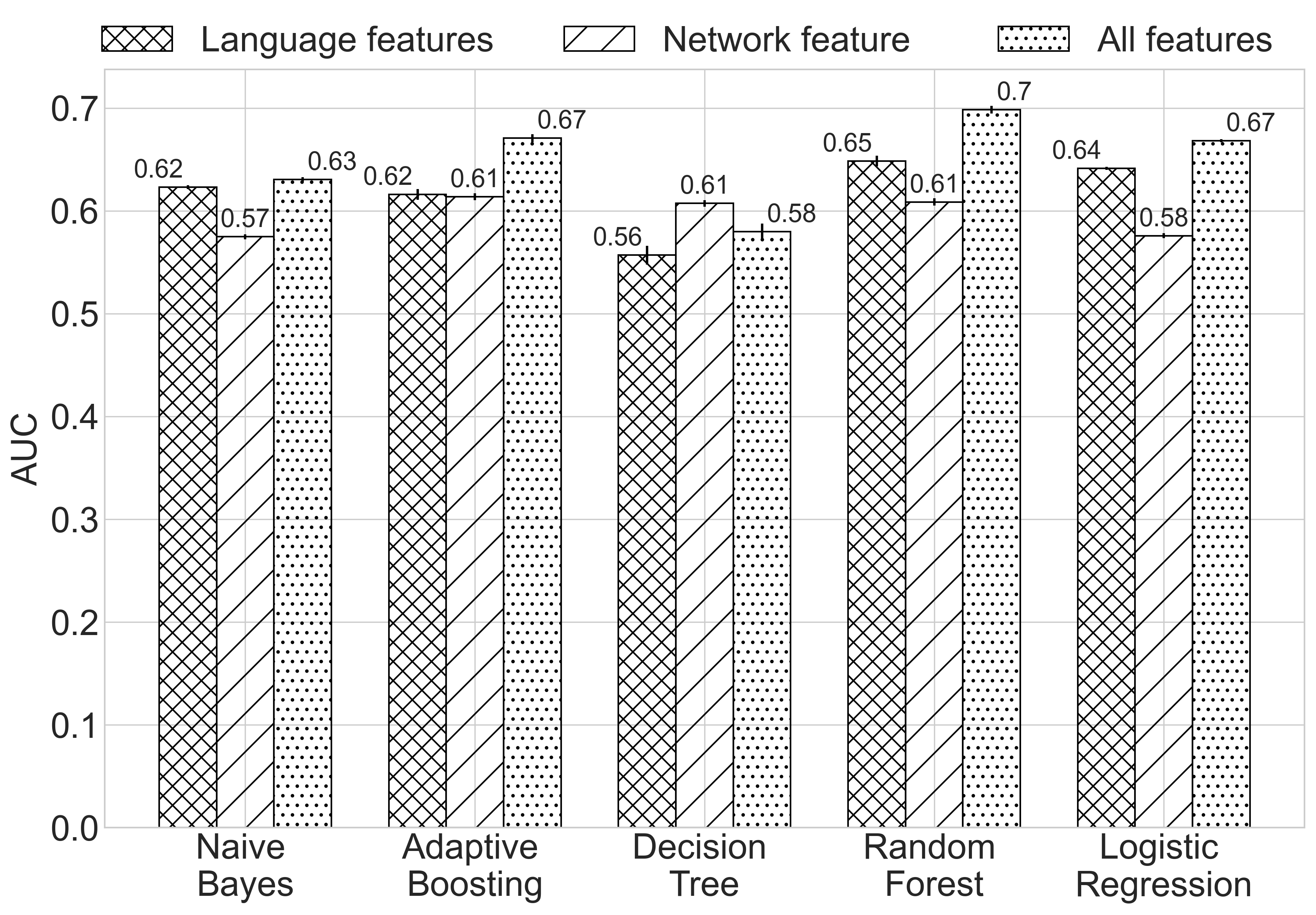}
    \caption{Area under the receiver operating characteristic curve (AUC) for language features, the network feature (degree ratio), and all features in each supervised learning model.}
    \label{fig:auc}
\end{figure}

We train and evaluate the performance of different supervised learning models based on language and/or the most significant network feature (degree ratio) and report the AUC for each feature group in Figure~\ref{fig:auc}. In support of the Elaboration Likelihood Model (ELM), language features are core to predicting persuasiveness. Still, we observe significant boosts in AUC from adding network features to models based on language features. Across the different classifiers, Random Forests attain the highest performance improvement (7.95\%) when adding network to language features. These findings suggest that network position plays a significant role in persuasion success. Across the different classification models, the Random Forest model performs best on all features (AUC=0.706).

\subsection{Persuasive arguments re-enforce the challengers' influential network position} When comparing the network position of challengers who received a $\Delta$ to their matches, we observe that receiving a $\Delta$ has non-trivial effects on successful challengers' network position (Table~\ref{tab:did-orders}).

\paragraph{In-degree.} We find that receiving a $\Delta$ increases the number of comments that a successful challenger receives, thus shifting other challengers' attention to the successful challenger ($\beta_3 = 2.219, p < 0.001$). 

\paragraph{Out-degree.} Receiving a $\Delta$ also increased a successful challenger's outgoing replies to others ($\beta_3 = 0.889, p < 0.001$). However, it is worth noting that receiving a $\Delta$ has a much higher effect on successful challengers' incoming than outgoing replies and is thus negatively associated with the degree ratio i.e., out-degree/in-degree ($\beta_3 = -0.618, p < 0.001$).

\paragraph{Authority.} We further observe that receiving a $\Delta$ increases the number of replies to a successful challenger from other challengers with high authority ($\beta_3 = 0.035, p < 0.001$). Thus, challengers who receive $Delta$s from OPs increasingly become important sources of information that are highly visible and influential to other challengers involved in the conversation.

\paragraph{Hubbiness.} Complementary to a challenger's importance as a source of information, we observe that receiving a $\Delta$ increases a successful challenger's hubbiness which represents their importance as gateways or connectors to other important challengers in the network thereby helping other challengers to navigate different arguments and viewpoints in the conversations ($\beta_3 = 0.065, p < 0.001$).

\paragraph{Betweenness centrality.} Additionally, we observe that receiving a $\Delta$ is associated with successful challengers occupy unique network positions and serve as bridges or intermediaries between different parts of the network ($\beta_3 = 112.061, p < 0.001$). Thus, successful challengers are better able to control or mediate the flow of ideas and viewpoints to other challengers during the discussions. 

\begin{table}[!h]
\resizebox{\columnwidth}{!}{%
\begin{tabular}{lll}
\hline
\begin{tabular}[c]{@{}l@{}}Network\\ Centrality\end{tabular} & Main analysis & Robustness check \\ \hline
In-degree & \begin{tabular}[c]{@{}l@{}}2.219***\\ (0.203)\end{tabular} & \begin{tabular}[c]{@{}l@{}}2.165*** \\ 0.204)\end{tabular} \\
Out-degree & \begin{tabular}[c]{@{}l@{}}0.889***\\ (0.157)\end{tabular} & \begin{tabular}[c]{@{}l@{}}0.588***\\ (0.156)\end{tabular} \\
Authority & \begin{tabular}[c]{@{}l@{}}0.035***\\ (0.006)\end{tabular} & \begin{tabular}[c]{@{}l@{}}0.040***\\ (0.007)\end{tabular} \\
Hub & \begin{tabular}[c]{@{}l@{}}0.065***\\ (0.014)\end{tabular} & \begin{tabular}[c]{@{}l@{}}0.006\\ (0.014)\end{tabular} \\
Betweenness & \begin{tabular}[c]{@{}l@{}}112.061***\\ (27.118)\end{tabular} & \begin{tabular}[c]{@{}l@{}}106.034***\\ (26.890)\end{tabular} \\ \hline
Observations & 7940 & 7786 \\ \hline
\end{tabular}%
}
\caption{Difference in Difference (DID) results on weighted communication networks. The main analysis includes replies to the winning argument and all subsequent arguments. The robustness check excludes replies to the winning argument.}
\label{tab:did-orders}
\end{table}

\subsection{Robustness checks}

To ensure the robustness of our findings, we conducted several additional analyses. First, we varied the evaluation metrics for our language and network feature-based classification models and obtained similar results across different measures, including accuracy and F1 score. Then, we used different selections of network features (e.g., in/out-degree, betweenness, hub, and authority) to complement language-based Random Forest classifiers and still observed significant improvements in prediction performance in terms of accuracy, F1, and AUC scores. However, we observe the highest performance improvement from degree ratio since it has the highest variance explained. Additionally, we examined the impact of $\Delta$ points on network position using both weighted and unweighted networks (i.e., considering the frequency as opposed to the mere presence of replies), and found consistent patterns of change in both cases. Furthermore, we conducted an analysis in which we ignored responses to the argument that won a $\Delta$, and observed similar changes in the network structure, indicating that successful challengers not only receive more interactions in response to their winning arguments but also attract greater attention from other participants in the conversation (Table~\ref{tab:did-orders}). Future research could extend these analyses to consider additional robustness checks, such as variations in network sampling methods or alternative model specifications, to further assess the conditions under which winning arguments benefit from the challengers' network position and the network position is further elevated after the recognition of a successful argument.

\section{Discussion}

It is widely accepted that we live in a connected world where people are often influenced by the attitudes, beliefs, and behaviors of others around them~\cite{easley2010networks,christakis2009connected}. Such social influences are prevalent in online social networks where people come together to share ideas, exchange information, and sometimes engage in debates. Our work focuses on the latter, where individuals engage in discussions in an attempt to persuade someone  to change their views on specific topics. Through our analysis, we sought to understand the dynamics of persuasion in this particular social context.

First, \textit{we investigated what makes some people successful persuaders}. Consistent with existing findings, we find that language features play an essential role in persuasion~\cite{tan2016winning,khazaei2017writing}. The language features included the use of external URLs to provide evidence in support of one's claims as well as the lexical diversity of one's arguments. These language features are important because they reflect the true merits of an argument. Our results align with the Elaboration Likelihood Model (ELM) theory, which suggests a central route that people use to process and evaluate the quality of arguments when making decisions~\cite{petty1986elaboration}. Therefore, our study emphasizes the crucial role of language and communication pathways in successful persuasion.

Second, \textit{we deduced emergent influence networks of who replies to whose arguments}, computed relevant network centrality measures, and quantified their role in persuasion. We found that adding network-based features to language-based supervised learning models significantly enhanced their predictive performance in classifying (un)successful challengers. Therefore, we believe that, although the emergent network position is not directly reflected in the arguments, it represent a peripheral cue. This finding suggest that people also rely on the peripheral route of information processing, e.g., by deducing heuristic cues about who to interact with based on their position in the influence network. Combined, these findings provide empirical support for the dual-processing nature of the ELM persuasion theory~\cite{petty1986elaboration}. 

Third, \textit{we investigate the effect of persuasiveness on network position} and find that successful persuasion leads to elevated centrality in the discussion network. Our difference-in-difference estimation shows that network centrality changes over time such that persuasiveness consolidates influential nodes' position in the network. For example, we observe that successful challengers benefit by more challengers interacting with them. Thus compared to unsuccessful challengers, successful challengers incur in-degree and authority score benefits which further enhances their influence in the network. This results is important in light of the pre-suasion model which suggests a way for challengers to influence others by capturing and channeling attention to themselves because it suggests that persuasuve challengers will capture others' attention and will be more likely to accrue more recognition~\cite{cialdini2016pre}.

Finally, \textit{we find that successful arguments have spillover effects.} We observe that successful challengers have more interactions with others, even when we ignore replies to the message that received a $\Delta$ in both weighted and unweighted networks. This means that successful challengers do not simply have more incoming interactions to their successful arguments, but also to other arguments they may have made. This observation also supports the pre-suasion model of persuasion in the sense that having a winning argument directs attention towards other arguments that one may have made, but were not publicly acknowledged as successful. Additionally, we observe that the degree distributions resemble a power-law distribution which tend to arises through a generative mechanism of preferential attachment~\cite{barabasi1999emergence}. Combined, these findings demonstrate reinforcing effect of successful arguments on network influence.

\paragraph{Broader impact.} We apply network science methods to provide a better understanding of persuasiveness in online discussions. We offer new insights on how to create effective communication strategies not simply based on the contents of the message, but also on the messenger's position in their social network. Our findings have broad implications for understanding the effectiveness of different messages and people's network positions in eliciting behavioral change to solve pressing problems around sustainability, migration, global health, etc. Thus, our findings are applicable to digital behavior interventions that aim to foster and support positive change~\cite{valente2012network}. For example, understanding what makes persuasion efforts successful can help overcome problems of low uptake or high attrition rates that are associated with behavior interventions in both online and offline settings. 

Our findings also have implications for democracy and political campaigns where emergent influence networks play a significant role in shaping political opinions and mobilizing voters. By understanding how network position relates to persuasiveness, researchers can gain a better understanding into the strategies used by political actors to persuade and mobilize their supporters, as well as the potential risks and opportunities associated with online political communication. 

\paragraph{Limitations and future work.} Since we do not conduct randomized controlled experiments, we cannot establish causal claims from the above findings. Our control group comprise a matched sample developed by \cite{tan2016winning} based on Jaccard similarity in vocabulary overlap between successful and unsuccessful arguments. While we maintain some consistency in terms of both data and methods to allow for comparisons with previous studies, we acknowledge that better state-of-the-art methods in Semantic Textual Similarity (STS) now exist, e.g., BERT~\cite{devlin2018bert} and RoBERTa~\cite{liu2019roberta}. Finally, our work focuses on persuasion in good-faith online discussions in a single Reddit community. We recognize that good-faith discussions are not overly representative of online spaces hence we do not know whether and how our findings may generalize to other forms of interaction that do not require mutual trust, honesty, and a willingness to engage in civil discourse, even when there are disagreements or differences in opinion.

\section{Ethics Statement}
As researchers, we recognize the potential ethical concerns that arise when using online user-generated content in research and have taken steps to address these concerns. Our study is based solely on publicly available data that has already been used in previous research, such as \cite{tan2016winning,khazaei2017writing,hidey2017analyzing,ji2018incorporating}. The raw data do not contain or reveal any sensitive information about users. To further protect the privacy of users, we have carefully considered the context in which the data are presented. We have presented only aggregated trends and do not reveal individual comments. We also acknowledge that our research findings may have social implications and therefore, to avoid any potential misapplication of our results, we take a neutral stance regarding the quality of arguments analyzed. Our focus is solely on the dynamics of persuasion and not on identifying who is right or wrong. Finally, we acknowledge the potential for manipulation and misinformation. The complexity of emergent influence networks can also create opportunities for manipulation and the spread of misinformation. By understanding the mechanisms of persuasion within these networks, we hope that researchers can develop strategies for countering the negative effects of misinformation and promoting accurate information online. While minimizing potential risks, we believe that the expected benefits of our contributions are substantial and outweigh unlikely and unintended harms.

\section{Conclusion}

In this study, we set out to investigate the impact of emergent influence networks on persuasiveness in unfacilitated online discussions. Through the use of a novel combination of social network analysis and machine learning, we were able to measure the influence of network position on persuasiveness, and demonstrate the impact of persuasiveness on successful users' network centralities over time. Our findings provide empirical support for the Elaboration Likelihood Model (ELM) of persuasion, and offer important insights into the complex social dynamics of online discourse. Looking ahead, our framework and methods could be applied in a variety of real-world settings to help organizations and individuals better understand how network position and persuasive strategies interact in digital spaces.


\bibliographystyle{unsrt}  
\bibliography{references}  

\end{document}